\documentclass[preprint,11pt]{elsarticle}

\usepackage{graphicx}
\usepackage{epstopdf}
\usepackage{subfigure}
\usepackage{array}
\usepackage{booktabs}
\usepackage{csvsimple}
\usepackage{pdflscape}
\usepackage{longtable}
\newcommand{\PreserveBackslash}[1]{\let\temp=\\#1\let\\=\temp}
\newcolumntype{C}[1]{>{\PreserveBackslash\centering}p{#1}}
\newcolumntype{R}[1]{>{\PreserveBackslash\raggedleft}p{#1}}
\newcolumntype{L}[1]{>{\PreserveBackslash\raggedright}p{#1}}

\usepackage{float}

\usepackage{amssymb}
\usepackage{lineno,hyperref}
\modulolinenumbers[5]
\graphicspath{{images/}}

\renewcommand{\figurename}{Fig.}
\graphicspath{{figures/}{materials/}}

\renewcommand{\figurename}{{\bf Fig.}}

\begin{document}
\begin{frontmatter}


\title{Optimization of the final settings for the Space-borne Hard X-ray Compton Polarimeter POLAR }
\author[h0] {Hualin Xiao \corref{xiaohlcor}}
\ead{hualin.xiao@psi.ch}
\author[h0]{Wojtek Hajdas\corref{wojtekcor}}
\author[h0]{Radaslow Marcinkowski}
\author[hp]{On behalf of POLAR collaboration}
\address[h0]{Paul Scherrer Institut, 5232 Villigen PSI, Switzerland }

\begin{abstract}
POLAR is a compact wide field space-borne detector dedicated for precise measurements of the linear polarization of hard 
X-rays emitted by transient sources in the energy range from 50 keV to 500 keV. 
It consists of 1600 plastic scintillator bars grouped in 25 detector modules that are used as gamma-ray detection material. 
Its energy range sensitivity is optimized for detection of the prompt emission photons from the gamma-ray bursts. 
Measurements of the GRB polarization provide unique information on emission mechanisms as 
well as on composition and structure of the GRB jets.
The POLAR instrument was developed by international collaboration of Switzerland, China and Poland. 
It was launched in space on-board the China Space Laboratory TG-2 on September 15th, 2016. 
Based on the ground calibration data, several high voltage and threshold settings were calculated and verified 
in order to obtain various energy ranges and optimized signal to background conditions for different 
measurement purposes. In this paper we present optimization procedure details and current test results.
\end{abstract}
\begin{keyword}
Gamma-ray Burst; Polarization;  Data analysis; Setting optimization;
\end{keyword}
\end{frontmatter}

\section{Introduction}

\begin{figure*}[htb]
\begin{minipage}{0.49\linewidth}
\includegraphics[width=0.95\textwidth]{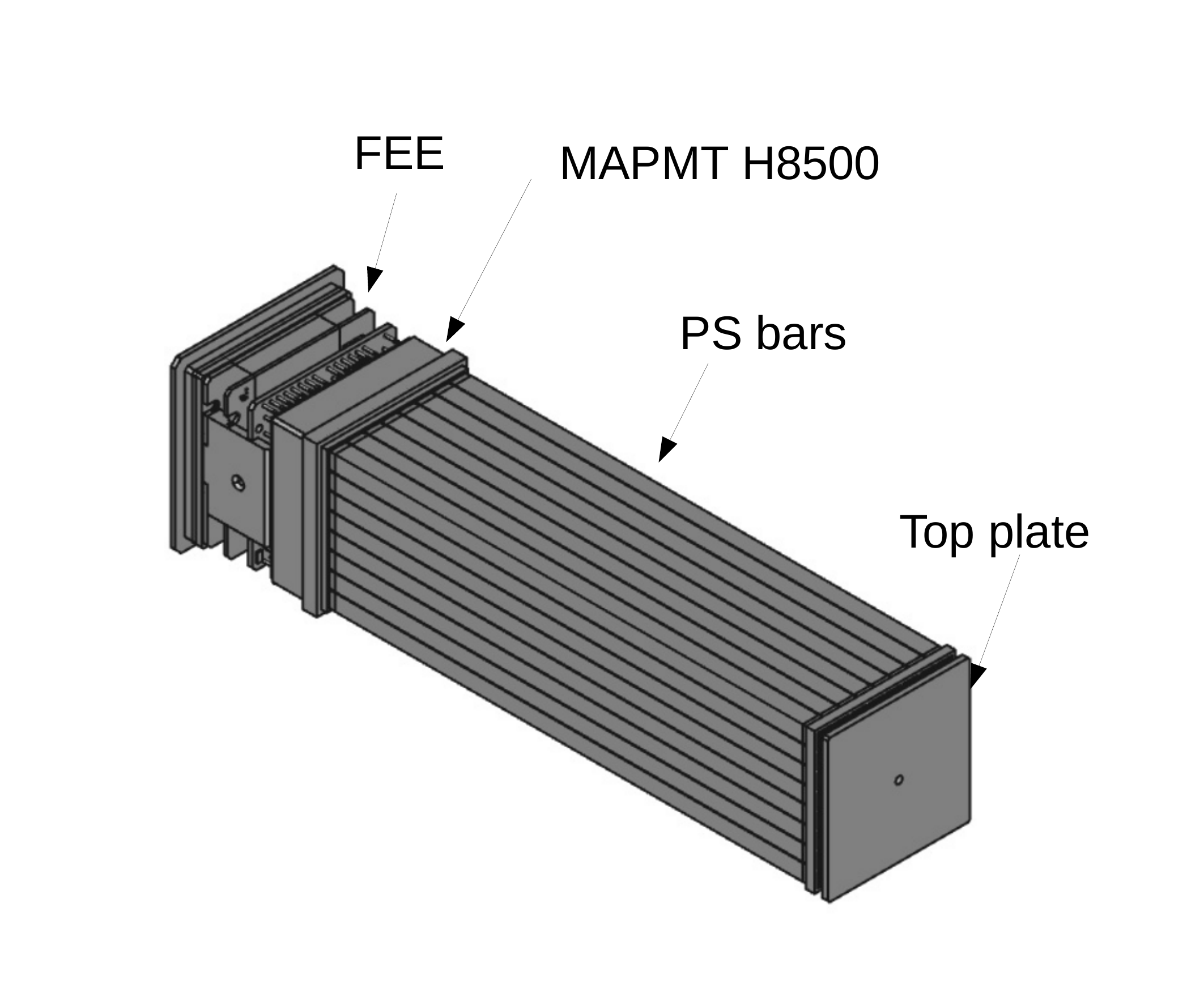}
\end{minipage}
\hspace{\fill}
\begin{minipage}{0.5\linewidth}
\includegraphics[width=0.95\textwidth]{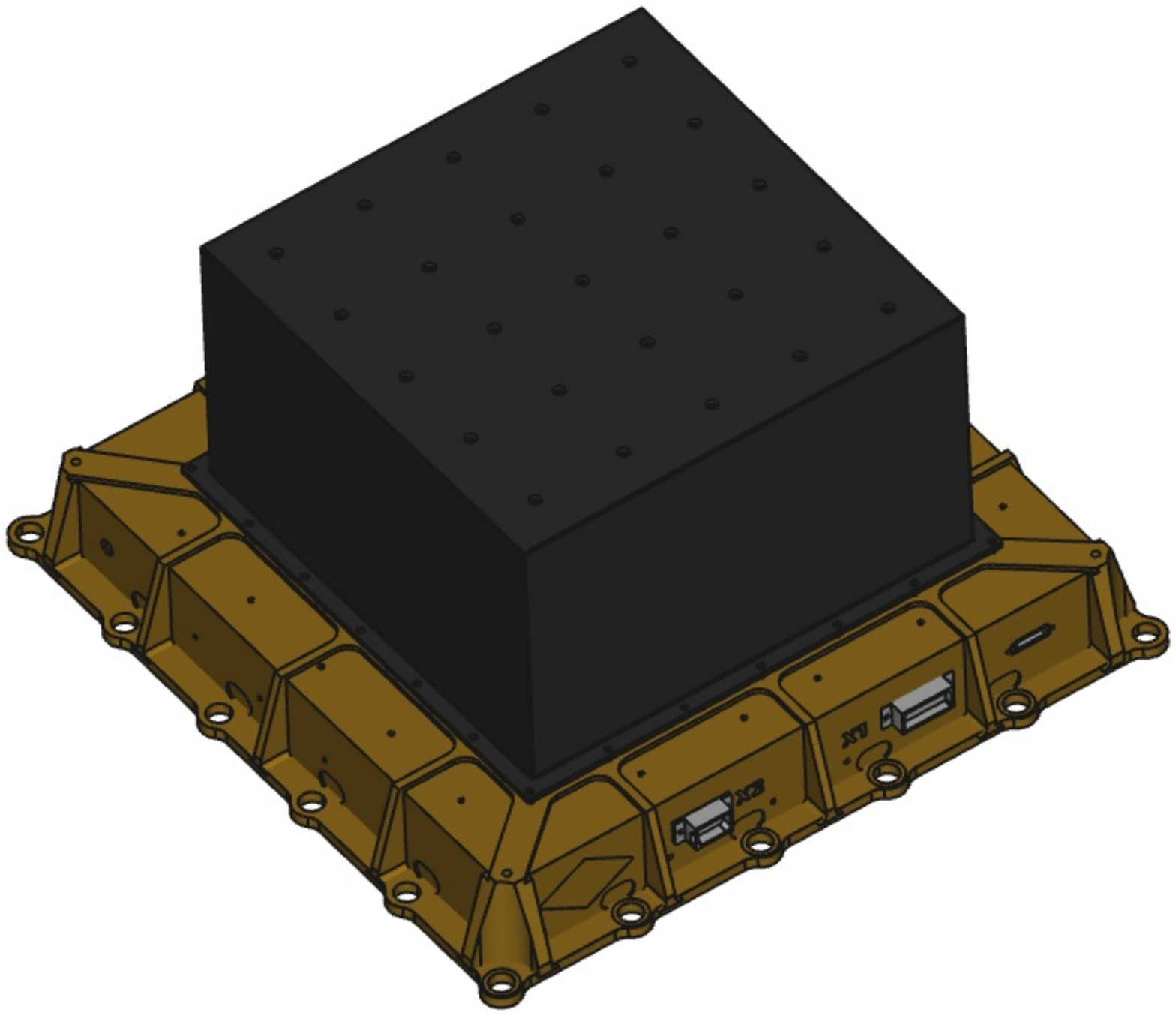}
\end{minipage}
\caption{POLAR detector module structure (left) and POLAR flight model (right).
Each module has 64 PS bars ($5.8 \times 5.8 \times 176\ \mathrm{mm}^3$ each) readout by a 64 channel MAPMT
(Hamamatsu H8500) 
and its  front-end electronics.
Note that the 1 mm thick carbon fibre socket  is not shown in the left panel.  
The full instrument consists of 25 identical modules.}
\label{fig:polar}
\end{figure*}
GRBs are sudden flashes of gamma-rays which appear randomly in the sky and for a few seconds outshine all other gamma-ray sources.
They are produced at cosmological distances and are considered as the brightest events in the universe after the Big Bang.
In the past 40 years many instruments have performed extensive studies of GRBs but both their creation mechanisms
and their progenitors are still uncertain.
Direct measurements of the linear polarization in the gamma-ray band  are thought to have a great
potential to distinguish between different theoretical models by providing missing information needed to pin down 
emission  mechanisms as well as composition and geometry structure of the GRB jets \cite{zb2,zb1}.

POLAR is a compact wide-field space-borne Compton polarimeter dedicated for precise measurements of the linear polarization of hard 
X-rays emitted by transient sources in the energy range from 50 keV to 500 keV. 
POLAR has both, a large effective detection area ($\sim$ 80 cm$^2$) and a wide field of view ($\sim$ 1/3 of full sky). 
For stronger GRB events its minimal detectable polarization (MDP) may be better than 10\%.
Due to its wide field of view, the instrument is also capable of polarization measurements in solar flares.
POLAR instrument was built by a collaboration of institutes from Switzerland, China and Poland. 
It was launched into space on September 15th, 2016 on-board the Chinese Space Laboratory TG-2 with a goal 
to reach up to 3 years long observation period. 

POLAR detector consists of 1600 plastic scintillator bars grouped in 25 identical detector modules.
Each module consists of 8$\times$8 PS bars, a 64 pixel multi-anode PMT (Hamamatsu MAPMT H8500) 
and a readout front-end electronics as shown in the left panel of Fig.~\ref{fig:polar}. 
Each bar has a dimension of $5.8 \times 5.8 \times 176$ mm$^3$ and is wrapped with a high
reflective foil (ESR).  
All 64 bars  are coupled to the MAPMT via an optical pad. 
The  FEE consists of three stacked Printed Circuit Boards (PCBs): high voltage
divider board, signal processing board and power supply and interfacing board.
The high voltage divider consists of twelve 470 k$\Omega$ resistors distributing high
voltages to the MAPMT dynodes. 
The signal processing board consists of a ASCI chip (IDEAS VA64) with 64 separate readout channels, ADC, DAC, 
FPGA, and a temperature sensor. The third board has a low voltage supply circuit and power and signal connecting interface.  
All 64 channels share the same high voltage divider; while each module has an individual high voltage supply.
The discriminators in the ASIC have a common threshold voltage value apart of small range trimmers.
Each detector module is packed with a 1 mm thick carbon fibre socket. 
All 25 modules are managed by the Central Task Processing Unit (CT). 
In addition, CT also manages power supplies, makes decisions on triggers and handles internal and external communication.
All above subsystems together with power supplies and interfacing parts are enclosed in a Carbon box.
It is shown  in the right panel of Fig.~\ref{fig:polar}. The whole instrument is mounted on the outside panel of the space-lab.

The main goal of POLAR is realized  by measuring the anisotropy of the azimuthal Compton scattering 
as described in the Klein-Nishina equation:
\begin{equation}
\frac{\mathrm{d}\sigma}{d\Omega}=
\frac{r_\mathrm{e}^2}{2}\left( \frac{E^{'}}{E} \right )^2
\left( \frac{E^{'}}{E}+\frac{E}{E^{'}}- 2 \sin{\theta}^2 \cos{\eta}^2 \right),
\end{equation}
where $r_\mathrm{e}$ is the classical radius of the electron, $E$ and $E^{'}$ are the energy of the incident photon and
the scattered photon, respectively, $\theta$ is the scattering angle between initial and final photon direction, 
and $\eta$ is the azimuthal scattering angle between the initial polarization vector and the direction of the scattered photon.
The  distribution of $\eta$, also called modulation curve,  
can be parametrized as:
\begin{equation}
 f(\eta)=k \cdot \left \{  1+ \mu \cos \left(2\left(\eta-\theta)+\pi\right)\right) \right\},
\end{equation}
where  $\theta$ is the polarization angle, 
$k$ is the normalization factor and $\mu$ is the modulation factor which is proportional to the polarization degree. 
The polarization degree can be acquired from $p=\mu/\mu_{100}$, where $\mu_{100}$ is the modulation factor
for 100\% polarized beam. 
Detailed descriptions of the POLAR principle and the polarization reconstruction methods can be found in Refs. \cite{nicolas, silvio, hlastro}.

In-flight energy calibration is carried out with four weak $^{22}$Na positron sources placed inside of POLAR.
As the sources emit two back-to-back annihilation photons, the selection of geometrically aligned 
coincidence events allows for proper calibration of each channel in the instrument. 
Energy calibration with other high voltage settings can be obtained using scaled energy conversion factors 
and experimentally determined high voltage dependence.
The annihilation gamma-ray pairs from calibration sources produce much higher energy depositions in scintillator bars
than most of gamma-rays from GRBs. Therefore, several high voltage and threshold settings are needed for comprehensive 
in-flight energy calibration and proper GRB observations.  
Before launching of POLAR, the high voltage vs. gain relation as well as threshold values were calibrated and several optimized settings for various measurement types were provided.
This paper presents  details of POLAR calibrations and describes optimization methods for its final settings.

\section{Calibration of POLAR flight model}
\subsection{Energy calibration}
\begin{figure}[htb]
\begin{center}
\includegraphics[width=0.6\textwidth]{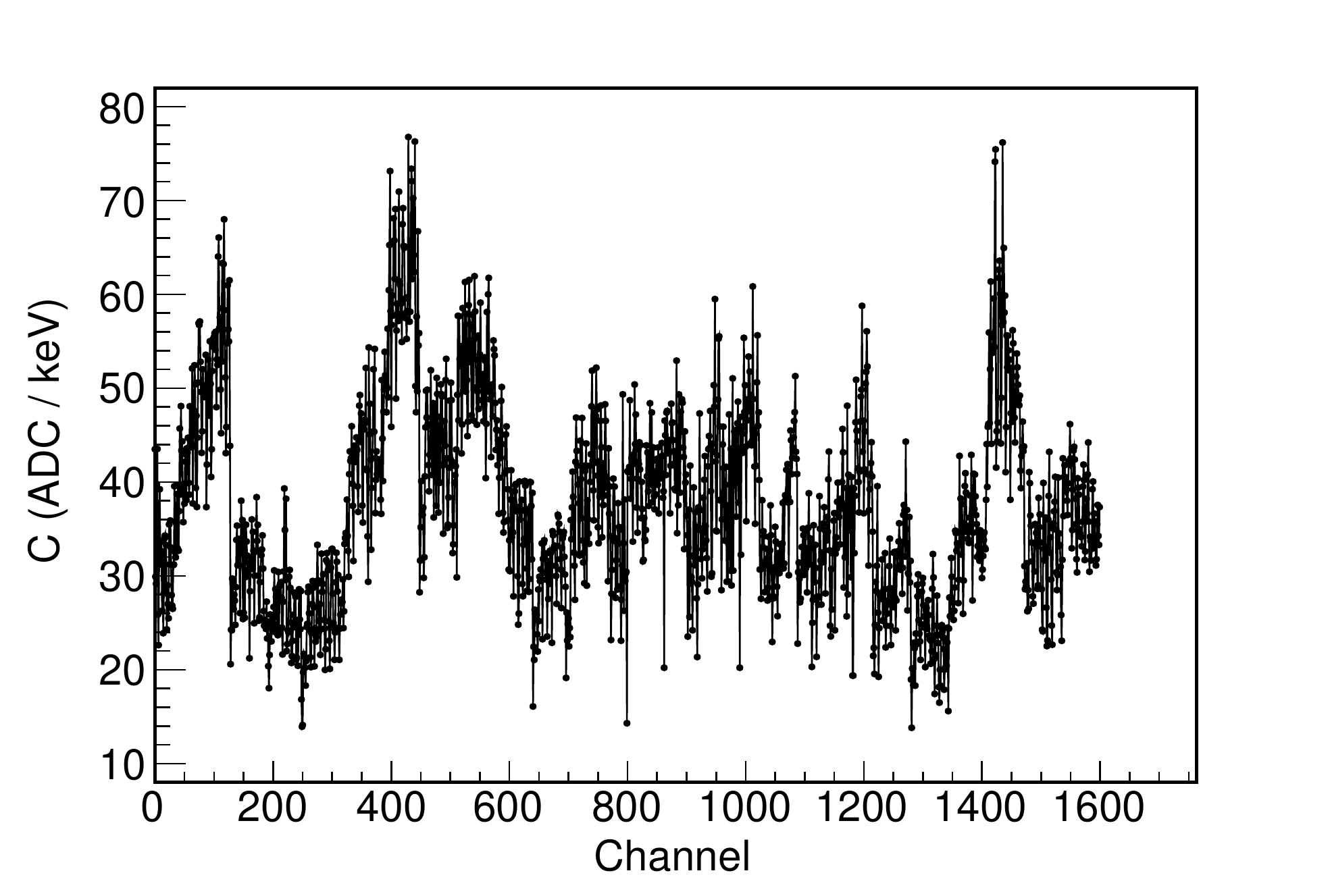}
\caption{Calculated energy conversion factors of all 1600 channels using Eq. (\ref{eq:cal}) parameterized in laboratory 
calibrations with $^{137}$Cs gamma-ray source. High voltage values of all modules were set to 700 V. 
}
\label{fig:cfactor}
\end{center}
\end{figure}

MAPMTs of POLAR operate using an equally distributed voltage divider. 
Their gain factor in relation to applied high voltage $V$ is
given by the following equation \cite{pmt}:
\begin{equation}
 G= a^n(\frac{V}{n+1})^{kn},
 \label{eq:gain}
\end{equation}
where $a$ is a constant, $n$ is the number of dynode stages and
$k$ is a constant determined by the structure and material of
the PMT. The typical value of $k$ is 0.7 \cite{pmt}.
In the case of POLAR, $n$ is equal to 12. 
According to Eq. (\ref{eq:gain}), the gain $G$ is proportional to the kn-th power of the high voltage.
It is reasonable to assume that the energy response of POLAR detector close to its typical operating condition is linear. 
The energy conversion $c$ (in units of ADC channel / keV) can be given by 
\begin{equation}
 c(V)=\frac{ E_{\rm meas}}{E_{\rm vis}}=b G=\alpha (\frac{V}{n+1})^{kn},
 \label{eq:cal}
\end{equation}
where $b$ is a constant, $V$ is the high voltage, $\alpha$ is equal to $b a^n$, 
$E_{\rm vis}$ and $E_{\rm meas}$  are the energy deposition  (in units of keV)
and the recorded energy deposition (in units of ADC channel) respectively. 
In order to determine the dependence between energy conversion factors and high voltage values, POLAR flight model 
(FM) was tested with the $^{137}$Cs source at six high voltage values from 610 V to 660 V with a step of 10 V. 
Fig. \ref{fig:cfactor} shows calculated energy conversion factors of all 1600 channels using Eq. (\ref{eq:cal}) parameterized with the calibration data. 
In this case the high voltage values of all modules were equal to 700 V.
More details about the method used in this study and the data analysis are available in Refs. \cite{hlieee} and \cite{xf}.

\subsection{Threshold calibration}
\begin{figure}[htb]
\begin{center}
\includegraphics[width=0.4\textwidth]{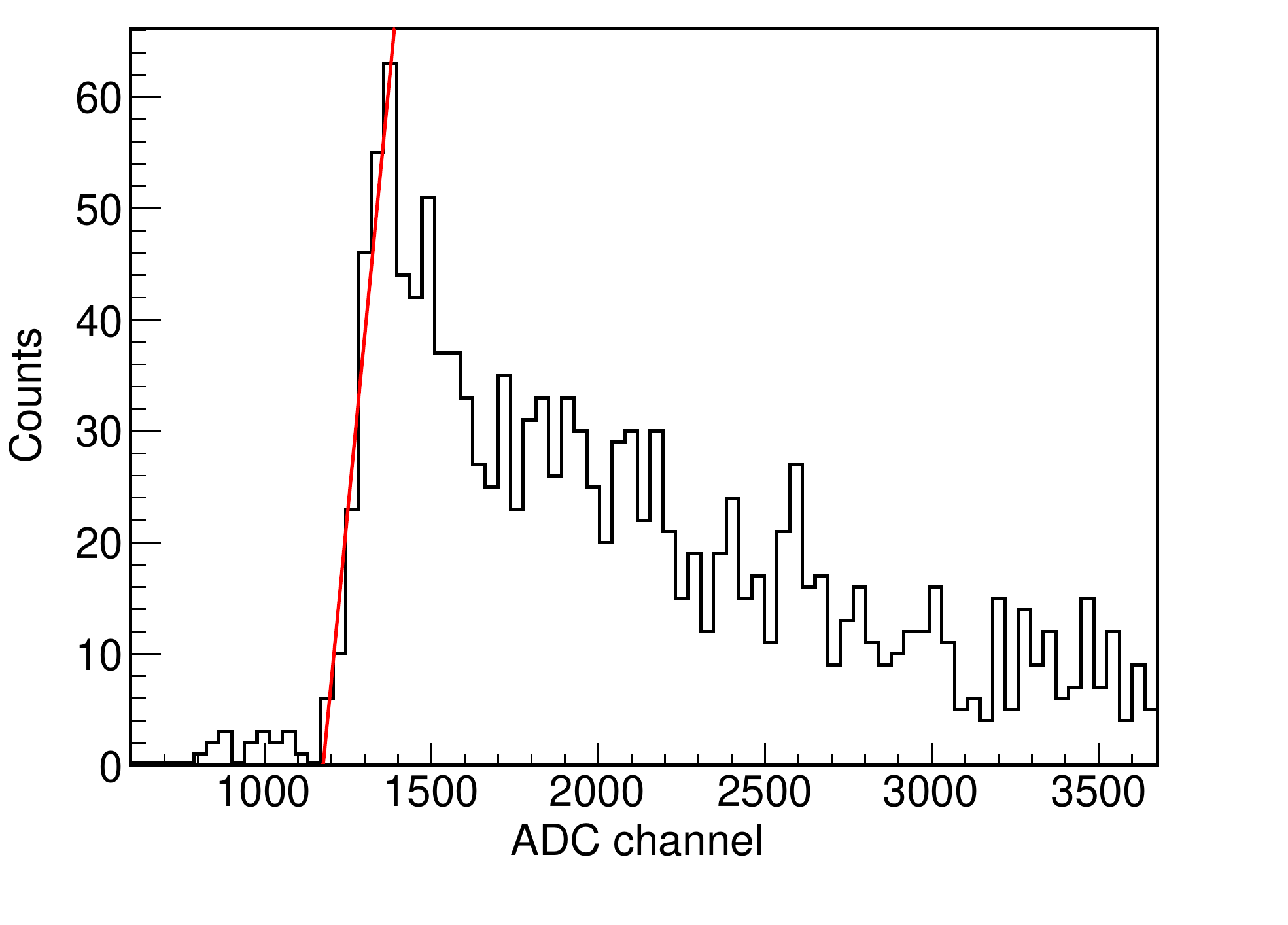}
\includegraphics[width=0.4\textwidth]{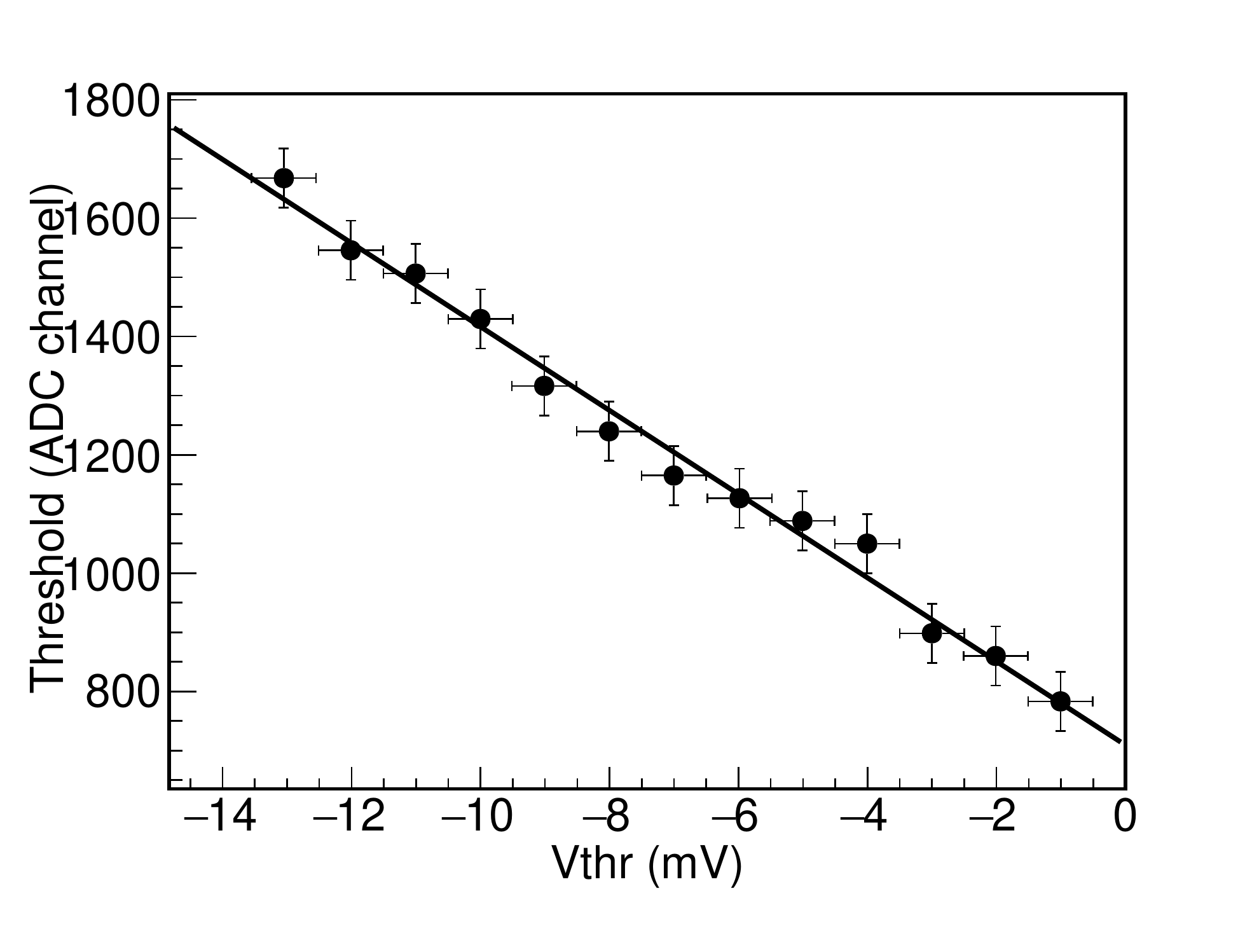}
\caption{ Example of the threshold position fit (left) and threshold positions  as a function of 
$V_{\rm thr}$ with a linear fit line (right).} 
\label{fig:thr}
\end{center}
\end{figure}

Precise knowledge of thresholds in all 1600 channels of POLAR is crucial as the value of its modulation factor computed for GRB polarization reconstruction is threshold dependent.
The ASICs in POLAR front-end electronics have a discriminator for each module with 64 channels. 
The discriminator can be set to the appropriate voltage value ($V_{\rm thr}$) of the threshold by sending commands
to its FEE.
The real threshold value $E_{\rm thr}$ in units of keV  can be given by 
\begin{equation}
E_{\rm th}=c \cdot (p_0+ p_1 V_{\rm thr}), 
\label{eq:thr}
\end{equation}
where $c$ is the
energy conversion factor, $p_0$ and $p_1$ are two constants  determined experimentally by 
studying the dependence between  threshold positions (in units of ADC channel)
and  $V_{\rm thr}$. In order to determine values of $p_0$ and $p_1$ in Eq. (\ref{eq:thr}),
we carried out several runs setting $V_{\rm thr}$ to different values and taking background data.
The left panel of Fig. \ref{fig:thr} shows an example of  the recorded energy spectrum for a pre-selected
channel for a typical background data run.  
As there are only a few hits amplitudes below the sharp cut-off on the left side of the spectrum we can assume that this effect is  caused by the hardware threshold.
The spectrum area around the cut-off was fitted with a line as shown in the left panel of Fig. \ref{fig:thr}.
The position of the half-maximum of the fitting range counts  gives the threshold position. 
The same fit procedure was repeated for all test runs.  
The right panel of Fig. \ref{fig:thr} shows threshold positions as a function of $V_{\rm thr}$. 
Values of $p_0$ and $p_1$ were obtained from a linear fit to the presented data points.

\section{POLAR final setting optimization}
\subsection{Optimization of settings for in-flight calibration}
\begin{figure}[htb]
\begin{center}
\includegraphics[width=0.4\textwidth]{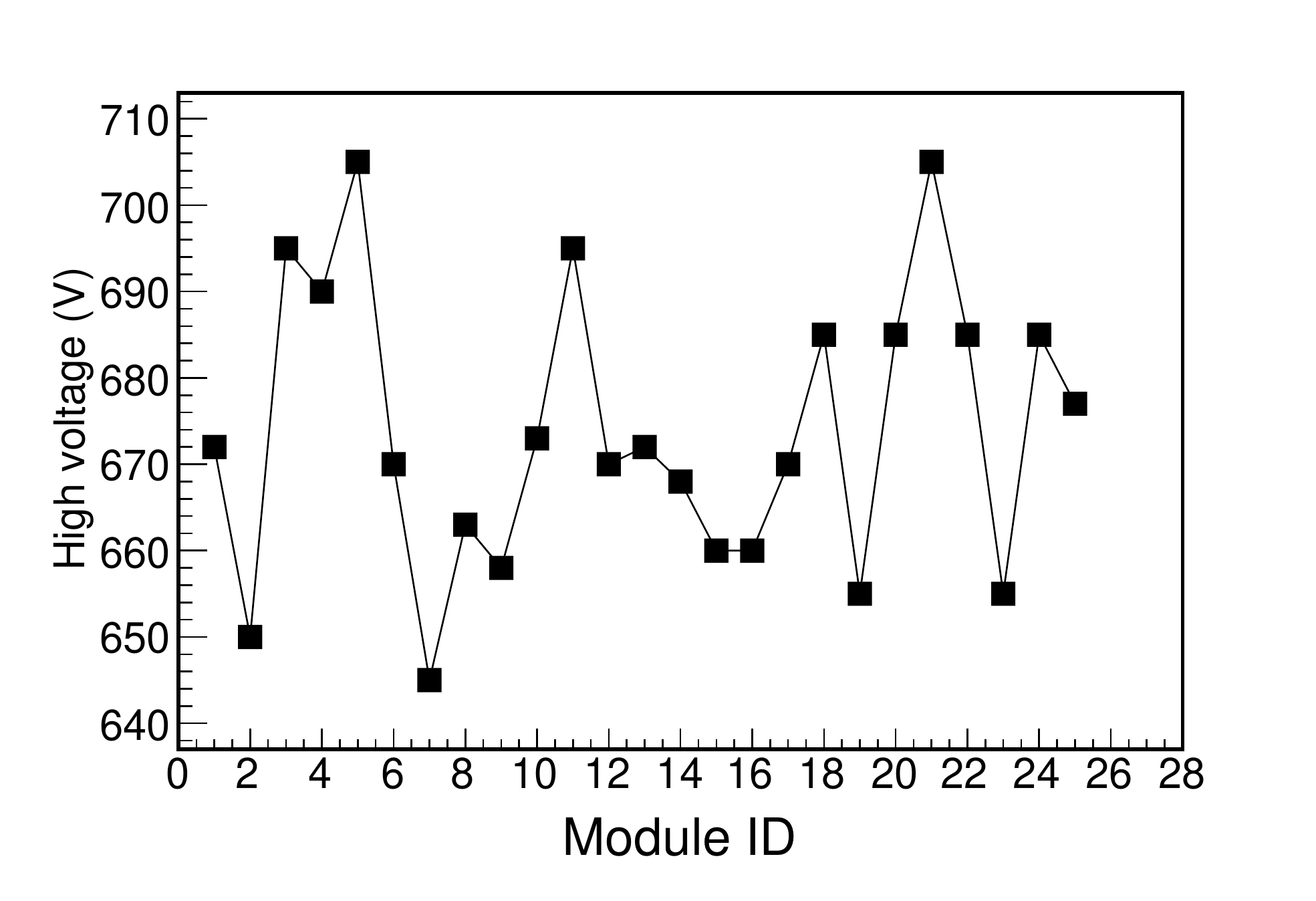}
\caption{POLAR basic high voltage setting for in-flight calibration.}
\label{fig:na22hv}
\end{center}
\end{figure}

POLAR has to undergo periodic calibrations during the mission in order to monitor energy response changes 
in its channels. The instrument gain factors may be susceptible to thermal drift, variations in high voltage values
as well as ageing and total dose effects. 
The in-flight calibration of POLAR is performed with four internal $^{22}$Na positron sources using positions of Compton edges in recorded energy spectra.  
The spectra are extracted with special conditions imposed on coincidence hits produced by x-rays from 
positron-electron annihilation.
The in-flight  calibration needs to be performed with several high voltage and threshold settings  
so that Eq. (\ref{eq:cal}) can be properly parametrized with such periodically determined conversion factors.
A proper high voltage and threshold setting for the in-flight calibration should meet  
the following requirements: 1) all Compton edges of 1600 channels (at 341 keV) measured with annihilation gamma-rays
are visible, i.e. the minimum detection range is around 400 keV assuming the energy resolution of about 20\%;
2) 25 modules have similar detection efficiencies for the 511 keV gamma-rays emitted by internal calibration sources;  
3) Threshold values of different modules are between 20 -- 60 keV and hot channels if any are ignored.
By using Eq. (\ref{eq:cal}) parametrized from the calibration with the $^{137}$Cs source in the laboratory, 
we calculated the high voltage setting meeting above requirements in all 25 modules. 
The high voltages are as shown in Fig. \ref{fig:na22hv}. 
The minimum energy range among 64 channels of each module is around 400 keV and 
the average expected Compton edge positions are at about 60\% of the ADC range.
This means that Compton edges should be seen in spectra of all 1600 channels. 
After knowing the high voltage values the threshold voltage ($V_{\rm thr}$) setting for each module was
optimized using Eq. (\ref{eq:thr}). 
The calculated average threshold value for 1600 channels is equal to about 40 keV. 
Due to gain non-uniformities in the modules the real, individual thresholds range from 30 keV to 60 keV.  

\begin{figure}
\begin{center}
\includegraphics[width=0.4\textwidth]{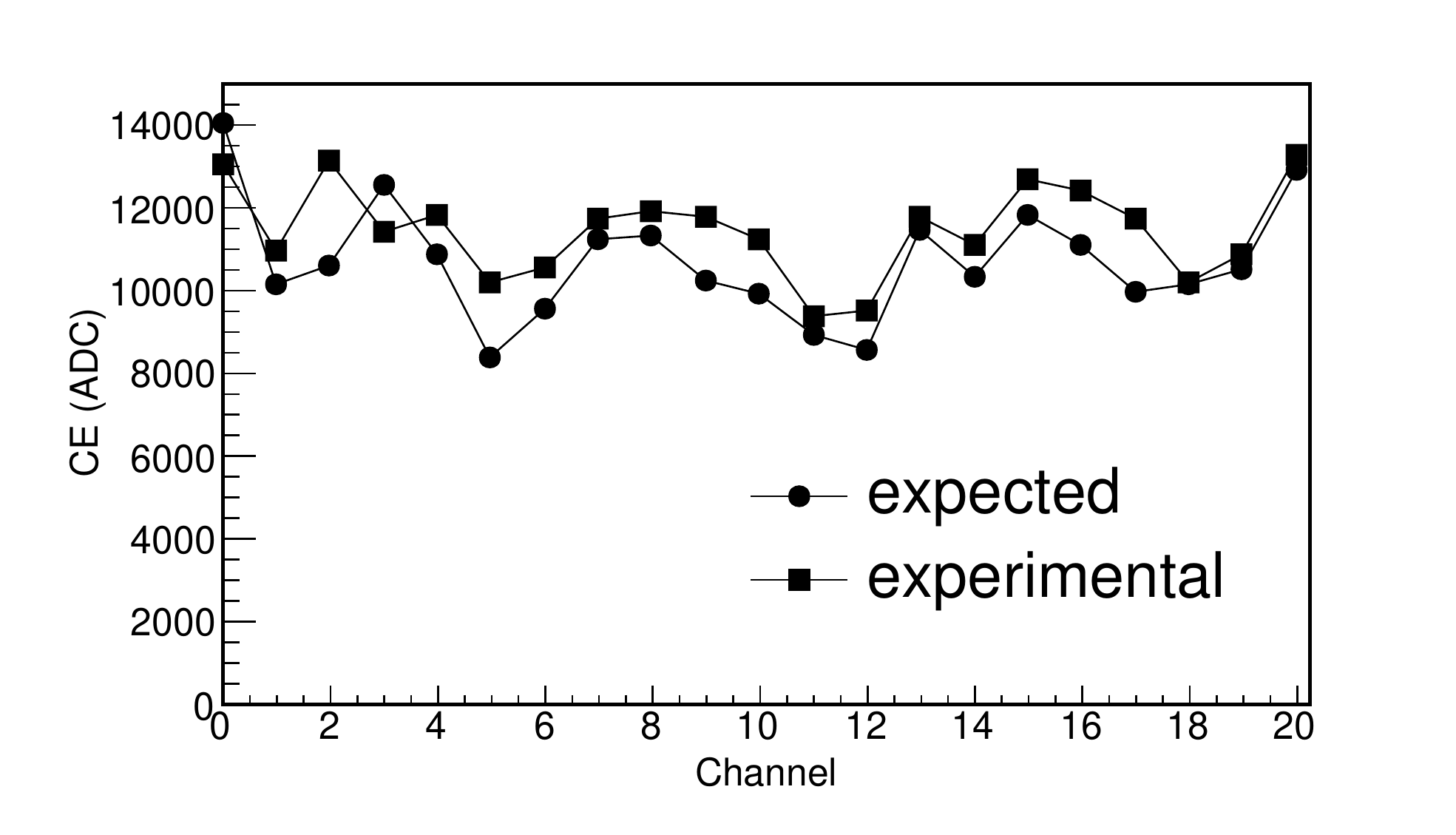} 
\caption{A Comparison between $^{137}$Cs experimental Compton edge positions  and calculations.}
\label{fig:cs137}
\end{center}
\end{figure}

In order to validate these calculations, POLAR FM was tested  with a 10 $\mu$ Ci $^{137}$Cs source in our laboratory.
The source was placed on the top of the FM about 15 cm above its cover. It was moved to the top of different modules 
during the test.
The Compton edges (at 477 keV) were also expected to be visible at least in energy spectra of the channels 
with lower gains. 
Fig. \ref{fig:cs137} shows an example of experimental and calculated Compton edge positions for 21 channels of POLAR before corrections 
for temperature effects. 
The mean temperature in POLAR FM modules during the latest test was equal to 25$^\circ$C.
It is about 10 degree lower than during previous laboratory calibrations aimed to determine   
energy conversion factors and high voltage dependence. In addition the temperatures of different modules 
were in general different. Taking it into account gives finally a good agreement between experimental 
and calculated positions of Compton edges.

The settings discussed above were stored in the flash memory of the POLAR FM to be used for in-flight calibration runs. 
In addition, we constructed and optimized another four settings for the same purpose. 
The high voltage values of each module differed by -21 V, -14 V, -7 V and +6 V relative to the 
values of the basic calibration setting. 

Several in-flight calibration runs of POLAR have already been performed after the launch using all five settings described above with typical results presented below. 
The first five panels of Fig. \ref{fig:inflight} show energy spectra for each of the settings constructed using 511 keV coincidence hits for POLAR channel 44.
The fit of the Compton edge  positions vs. the high voltage values is shown in the sixth panel of Fig.~\ref{fig:inflight}. 
Note that the ADC channel values in the space data packets were scaled down by the factor 1/4 in order to reduce size of the the data packet.

\begin{figure}
\begin{center}
\includegraphics[width=0.7\textwidth]{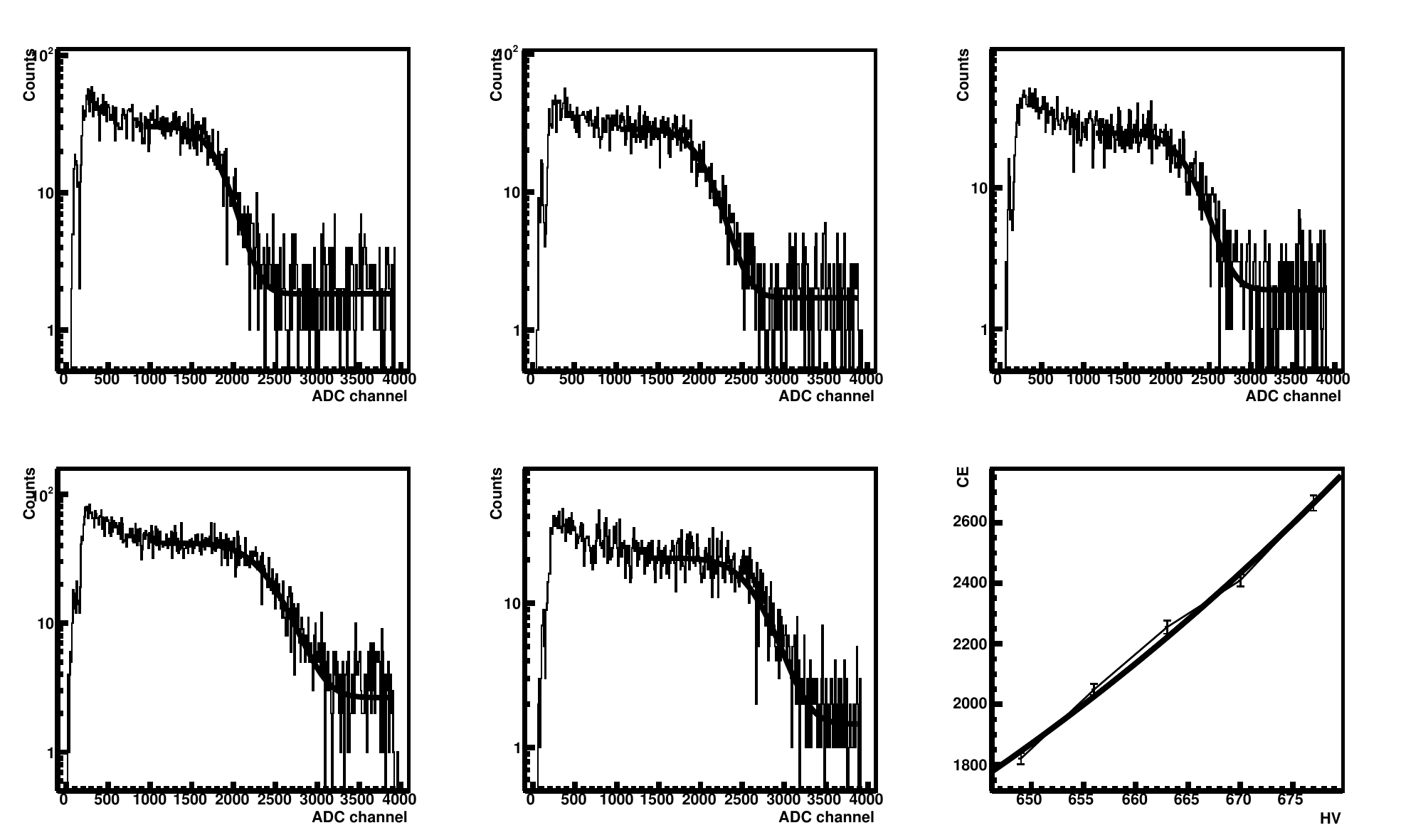} 
\caption{Example of the POLAR in-flight calibration data for channel 44. Five panels show energy spectra of coincidence hits at five different high 
voltage values; the last panel plot shows a fit of Compton edge positions vs. high voltage values. 
}
\label{fig:inflight}
\end{center}
\end{figure}

\subsection{Optimization of settings for GRB observation}

\begin{figure}
\begin{center}
\includegraphics[width=0.6\textwidth]{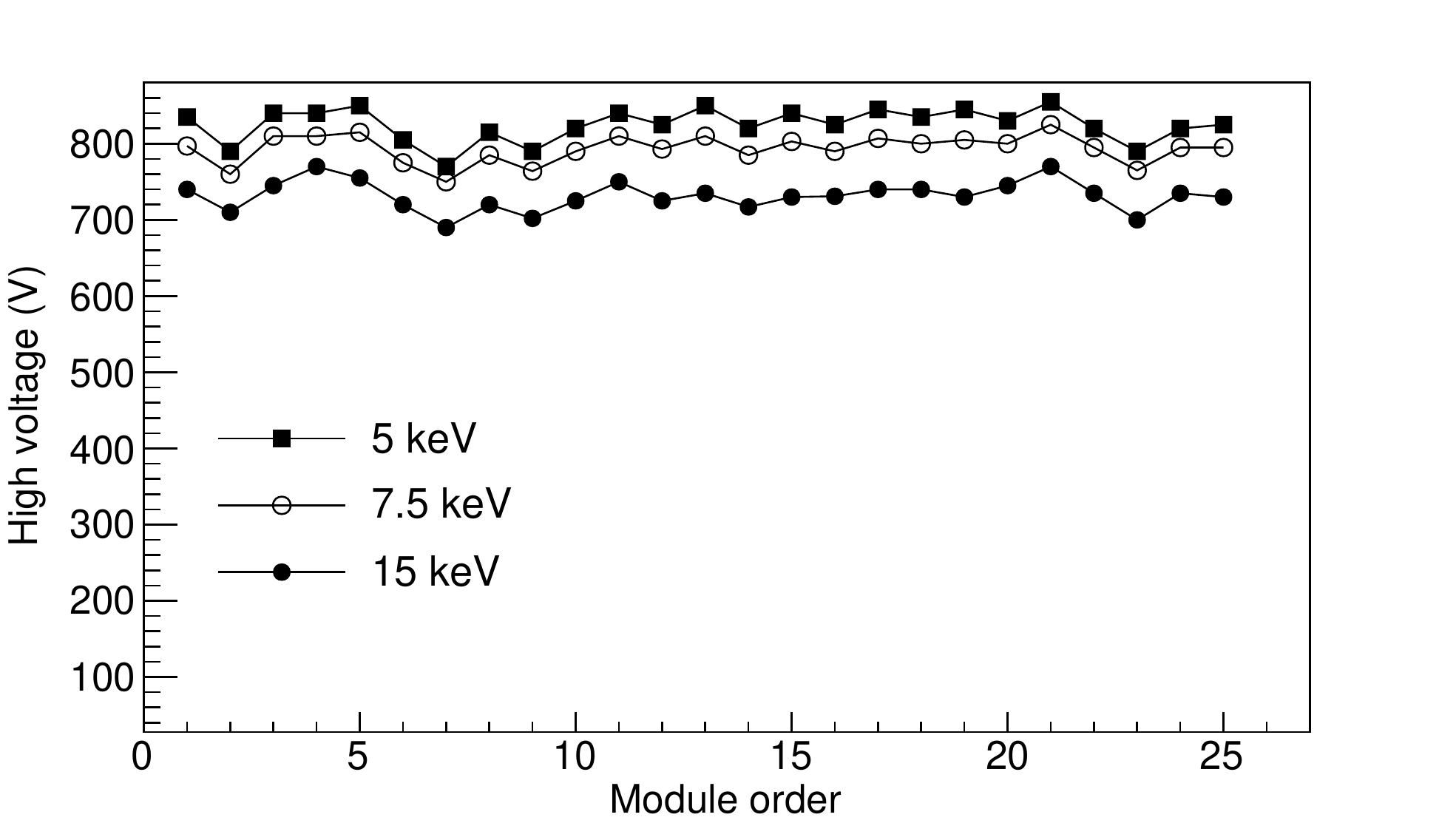} 
\caption{
MAPMT high voltage values of three settings optimized for GRB observation. 
The mean thresholds of the three settings are 15 keV, 7.5 keV and 5 keV.
}
\label{fig:grbhv}
\end{center}
\end{figure}

Anticipating various signal to background and noise conditions in space we calculated for each module three high voltage and discriminator threshold
settings using parametrized Eqs.~(\ref{eq:cal}) and (\ref{eq:thr}).
The mean values of thresholds according to calculations were set to energies equal to 15 keV, 7.5 keV and 5 keV.  Corresponding high voltage values are shown in Fig. \ref{fig:grbhv}.
In average all modules were expected to have very close values of energy conversion factors and also the thresholds 
so that detection efficiencies would be similar. 
Directly after applying these settings to POLAR FM several hot channels could be observed. 
Therefore the thresholds in corresponding modules were increased to reduce counting rates and numbers of accidental  coincidences caused purely by hot channels.
The background counting rates in the laboratory measured for these three final settings were about 220 Hz, 400 Hz and 500 Hz per module respectively. 
The measurements were performed in the laboratory with typical temperatures about 20$^\circ$C.
Further verification in space reveal very high levels of the low energy background. In order to reduce the background rate and provide enough resources in the telemetry bandwidth 
for GRB detections mostly the setting with the mean threshold value of 15 keV have been applied. 

\section{Conclusion}
POLAR is a compact wide-field of view, space-borne detector dedicated for precise measurements of the linear polarization of hard X-rays emitted by 
transient sources in the energy range from 50 keV to 500 keV.
Based on the parametrized energy conversion factor, their high voltage dependence and the calibrated values of thresholds, 
sever settings were constructed and optimized for in-flight calibrations and observations of GRBs. 
In-flight calibrations of POLAR in space were performed successfully for several times. Typical settings for 
routine operation utilise the one with mean threshold value of 15 keV. To date i.e. within six months of
operation in space more than 50 GRBs have been already detected.

\end{document}